\documentclass{emulateapj}
          
\newif\ifAMStwofonts                        
\newcommand{\lsimeq}{{_<\atop^{\sim}}}
\newcommand{\gsimeq}{{_>\atop^{\sim}}}

\newcommand{\mic}{$\mu$m~}

\shorttitle{Mid-Infrared Source Counts at 24-$\mu$m}
\shortauthors{Gruppioni et al.}

\begin{document}

\title{Extragalactic Source Counts in the {\em Spitzer} 24 Micron Band: 
What Do We Expect from ISOCAM 15 Micron Data and Models?}

\author{C. Gruppioni\altaffilmark{(1)}, F. Pozzi\altaffilmark{(2)}, 
C. Lari\altaffilmark{(3)}, S. Oliver\altaffilmark{(4)} and G. Rodighiero\altaffilmark{(5)} }

\altaffiltext{(1)}{Istituto Nazionale di Astrofisica: Osservatorio Astronomico di Bologna, via Ranzani 1, I--40127 Bologna, Italy. \\ e-mail: carlotta.gruppioni@bo.astro.it}
\altaffiltext{(2)}{Dipartimento di Astronomia, Universit\`a di Bologna, via Ranzani 1, I--40127 Bologna, Italy}
\altaffiltext{(3)}{Istituto di Radioastronomia del CNR, via Gobetti 101, I--40129 Bologna, Italy}
\altaffiltext{(4)}{Astronomy Centre, Department of Physics \& Astronomy, University of Sussex, Brighton BN1 9QJ, UK}
\altaffiltext{(5)}{Dipartimento di Astronomia, Universit\`a di Padova, vicolo dell'Osservatorio 2, I--35122 Padova, Italy}

\begin{abstract}
The comparison between the new {\em Spitzer} data at 24 $\mu$m and the previous
ISOCAM data at 15 \mic is a key tool to understand galaxy properties and evolution in the
infrared and to interpret the observed number counts, since the combination of {\em Spitzer} with
the {\em Infrared Space Observatory} cosmological surveys provides for the first
time the direct view of the universe in the infrared up to $z \gsimeq 2$.
We present the prediction in the {\em Spitzer} 24 $\mu$m band of a phenomenological 
model for galaxy evolution derived from the 15 $\mu$m data. 
Without any ``a posteriori'' update, the model predictions seem
to agree well with the recently published 24 $\mu$m extragalactic source 
counts, suggesting that the peak in the 24 \mic counts is dominated by starburst galaxies like 
those detected by ISOCAM at 15 $\mu$m but at higher redshifts 
($1 \lsimeq z \lsimeq 2$ instead of $0.5 \lsimeq z \lsimeq 1.5$).
\end{abstract}


\section{Introduction}

Cosmological constraints on the evolution of galaxies have been recently investigated
by studying the statistical properties of large samples. In particular, the combined
analysis of extragalactic source counts and redshift distributions is generally
used to calibrate the theoretical models as a function of time.
The mid-infrared (MIR) and far-infrared (FIR) regions of the electromagnetic spectrum
probe the population of actively star-forming galaxies obscured by dust.
Extragalactic source counts from different surveys over a wide flux range
obtained with the ISOCAM instrument on board of the {\em Infrared Space Observatory} ({\em ISO})
indicate that these sources have evolved rapidly, significantly faster than deduced from 
optical surveys (Elbaz et al. 1999; Gruppioni et al. 2002; Metcalfe et al. 2003; Rodighiero et al. 2004).  
These results are supported by the detection of a substantial cosmic infrared background 
(CIRB; Hauser \& Dwek 2001), which is interpreted as the integrated 
emission from dust present in galaxies. The contribution of resolved ISOCAM 
sources accounts for $\sim$60-70\%
of the CIRB at MIR frequencies, the bulk of this background originating in discrete sources
at $z \lsimeq 1.2$ (Franceschini et al. 2001; Elbaz et al. 2002).

\begin{figure}
\plotone{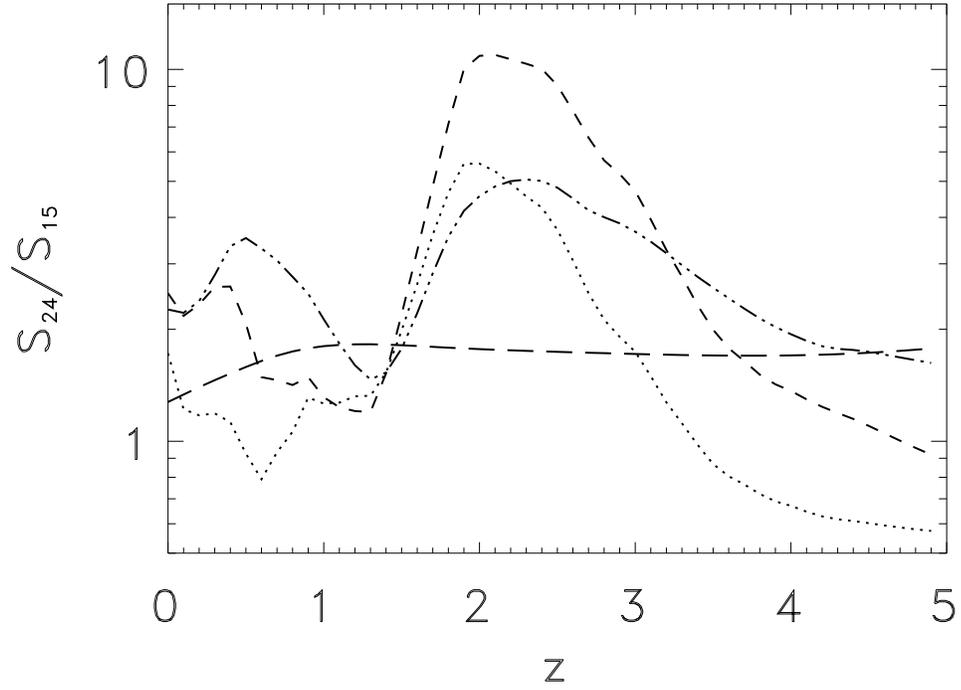}
\caption{$S_{24 \mu m} / S_{15 \mu m}$ ratio as a function of redshift for the MIR populations
contributing to the observed number counts:
starburst galaxies (short-dashed line: M82 SED prototype), normal galaxies (dotted line: M51), 
type 2 AGN (dot-dot-dot-dashed line: Circinus) and type 1 AGN
(long-dashed line: SED from Elvis et al. 1994). 
}
\label{fig1}
\end{figure}

The {\em Spitzer Space Telescope} is now providing new insights into the
IR population contributing to the CIRB, in particular with the Multiband Imaging Photometer 24 $\mu$m band, 
which is starting to detect a population of galaxies that may be IR-luminous galaxies at $z\sim 1.5-3$
(i.e. high-redshift analogs of the faint 15 \mic galaxies detected by ISOCAM; see
Chary et al. 2004).
Preliminary results from the 24 $\mu$m extragalactic source counts (Marleau et al. 2004;
Papovich et al. 2004), confirming the existence of the rapidly evolving dust-obscured population
discovered byISOCAM, raise the question 
on how to compare them with the previous ISOCAM counts at 15 $\mu$m. 
Both 24  and 15 $\mu$m bands are extremely sensitive to the presence of broad emission features at
6.2, 7.7, 8.6, 11.3 and 12.7 \mic in the spectra of galaxies,
probably from polycyclic aromatic hydrocarbons (PAHs; Puget \& Leger 1989). Since these features dominate the photometric
output at some redshifts, the ratio between the {\em Spitzer} 24 \mic 
and the ISOCAM 15 \mic fluxes ($S_{24 \mu m} / S_{15 \mu m}$)
is strongly dependent on $z$.
In figure \ref{fig1} this ratio versus $z$ is shown for the populations contributing to the 
MIR source counts (see next section for a discussion): 
starburst galaxies (M82 spectral energy distribution [SED]), normal
galaxies (M51), type 1 active galactic nuclei (AGNs; SED from Elvis et al. 1994), and 
type 2 AGNs (Circinus).
As clearly shown in the plot, 
the 15 \mic band is optimized for detecting $0.5 \lsimeq z \lsimeq 1.5$ galaxies,
while the 24 \mic band is favored for the detection of galaxies at
$z \gsimeq 1.5$.  
Since the comparison between the source counts in the two bands is a powerful tool
for understanding the evolutionary properties of the different galaxy populations contributing 
to the counts at different redshifts and flux levels, it is worthwhile performing a careful
comparison.

In a recent paper on extragalactic source counts from the First Look Survey (FLS), Marleau et al. (2004)
try to compare the 24 $\mu$m source counts with
the previous 15 $\mu$m counts from different ISOCAM surveys, transformed to
24 $\mu$m. However, the reported transformation appears to be overly simplistic, since
the ISOCAM counts (plotted in figure 4 of Marleau et al. (2004)) have been 
converted to the {\em Spitzer} 24 $\mu$m band by considering a single value for the
$S_{24 \mu m} / S_{15 \mu m}$ flux ratio for all flux densities. The value used by the authors
($S_{24 \mu m} / S_{15 \mu m}$ = 1.2) is a median value derived from typical luminous infrared
galaxy/ultraluminous infrared galaxy SEDs at 
relatively high $z$ (Chary \& Elbaz 2001), which is appropriate only for sources making up the
peak of the counts. In fact,
as shown in figure \ref{fig1}, if we consider the ``starburst'' template (M82), we obtain a local 
value for the flux ratio of $\sim$2.5, while
we reach a value of $\sim$1.3 only at $z \sim 1$. Therefore, only the contribution 
to the counts made by galaxies with $z$ around 1 could be transformed to 24 $\mu$m using a flux ratio 
similar to that considered by Marleau et al. (2004). In particular,
the bright part of the European Large-Area {\em ISO} Survey (ELAIS) counts (Gruppioni et al. 2002), linking the {\em IRAS} counts to the deep ISOCAM counts,
is dominated by nearby non-evolving normal galaxies(with ratios of $\sim$1.7) and by starburst galaxies 
and type 2 AGNs (with ratios of $\sim$2.5 and $\sim$2.3, respectively), as shown by La Franca et al. 
(2004). Therefore, the use of a single ratio 
value of $1.2$ produces a misleading result [especially at $S_{24 \mu m} > 1$ mJy: 
i.e. 15 $\mu$m counts
shifted downwards by a factor of $(1.7/1.2)^{1.5} - (2.5/1.2)^{1.5}
= 1.7 - 3.0$], suggesting an apparent inconsistency
between the bright part of the source counts in the two bands.

Since at the moment there are no large areas with available data at both wavelengths, we can make 
use of a model fitting the observed 15 \mic source counts (Pozzi et al. 2004, I. Matute et al. 2005, in preparation) to
transform the counts from one frequency to the other, thus allowing a direct comparison.

In this Letter we discuss the more realistic way to transform the model predictions and
the observed data from the LW3 band of ISOCAM to the 24 $\mu$m band of {\em Spitzer},
in order to compare the properties of the 24 \mic sources with those of the ISOCAM 15 \mic ones.
The Letter is structured as follows: in section \ref{model}, we describe the evolutionary
model fitting the 15 \mic observables; in section \ref{exp24mic}, we show the model predictions
at 24 $\mu$m; in section \ref{15mic_24mic}, we discuss a method to transform
the observed data points from 15 to 24 $\mu$m; in section \ref{concl}, we
present our conclusions.  

Throughout this Letter we will assume $H_0 = 75$ km s$^{-1}$ Mpc$^{-1}$, 
$\Omega_m = 0.3$ and $\Omega_{\Lambda} = 0.7$.

\section{The Model}
\label{model}
The model is based on the first direct determination of the 15 $\mu$m luminosity function (LF) 
of galaxies and AGNs, based on data from the ELAIS southern fields survey
(Lari et al. 2001; La Franca et al. 2004, Rowan-Robinson et al. 2004).  We assume that four
main populations, evolving independently, contribute to the observed source counts: starburst
and normal galaxies and type 1 and 2 AGN. A maximum likelihood analysis
(Marshall et al. 1983) has been used to simultaneously fit both
evolution rates and 
shape parameters of the different local LFs.
Although AGN make up a non negligible fraction of the extragalactic source counts at
15 $\mu$m (especially at high flux densities), galaxies are the dominant class in the
MIR.

\begin{figure}
\plotone{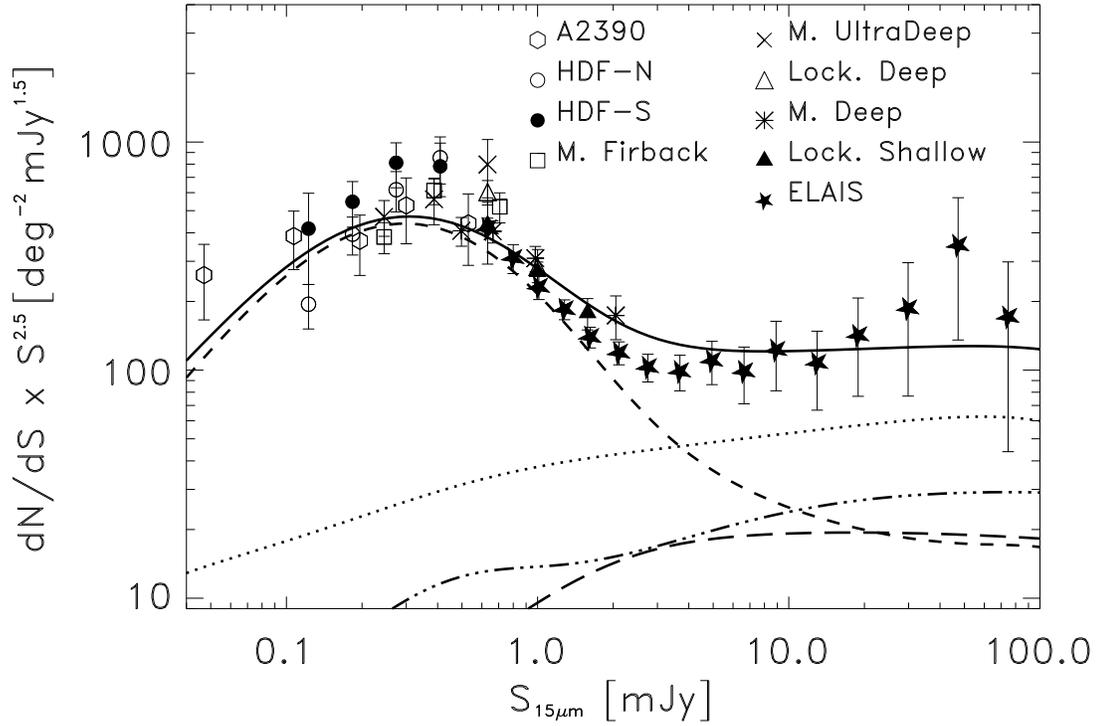}
\caption{Normalised differential source counts in the ISOCAM 15 \mic band.
As explained in the plot, data points are from several surveys (HDF-N,
HDF-S, Marano Firback, Ultra-Deep and Deep: Elbaz et al. 1999b;
Ultra-Deep lensed: Metcalfe
et al. 2003; ELAIS-S1: Gruppioni et al. 2002; Lockman Deep and
Shallow: Rodighiero et al. 04). The model curves
are from Pozzi et al. (2004) for galaxies (dash: starburst; dot: normal) and
from Matute et al. (in preparation) for AGN (long-dash: type 1; dot-dot-dot-dash: type 2).}
\label{fig2}
\end{figure}

The LF determination for galaxies, described extensively in Pozzi et al. (2004), is based  
on a sample of about 150 galaxies in the
redshift interval $0.0 \leq z \leq 0.4$, covering a large flux density range between 
{\em IRAS} and the deep ISOCAM surveys ($0.5 \leq S_{15 \mu m} \leq 50$ mJy). The normal,
non evolving, and the starburst, evolving, populations are
separated using the new criterion based on the MIR to
optical luminosity ratio ($L_{15 \mu m} / L_R$). 
We use the basic Silva et al. (1998)
models to reproduce the SED of our prototypical galaxies, assumed to be M82 for
the starburst population and M51 for the normal one. The MIR region (between 3 and 
18 $\mu$m) of the modeled spectra were replaced with ISOCAM circular variable filter observations
(M82: Forster-Schreiber et al. 2003; M51: Roussel et al. 2001).
Note that, for simplicity, we have used a single template SED for each
population, instead of different SEDs for different infrared luminosity intervals.
While the normal 
population is consistent with no evolution, for the starburst population a
strong evolution is found both in luminosity [$L(z) \propto (1+z)^{3.5}$ up to $z\sim1$]
and in density [$\rho(z) \propto (1+z)^{3.8}$ up to $z\sim1$]. 
The evolutionary parameters
of our model have been tested by comparing the model predictions with all the available
observables, like source counts at all flux density levels (from 0.1 to 300 mJy) and
redshift distributions and LF at high $z$. The agreement between the 
model predictions and the observed data is remarkably good (see figure \ref{fig2} for an 
example of how the model fits the observed number counts at 15 \mic). 

The LF determination for AGNs (both type 1 and 2), described in I. Matute et al. (2005, in preparation), 
is based on ELAIS data (27 type 1 AGNs and 25 type 2 AGNs) combined with the local {\em IRAS} sample at 12 $\mu$m of 
Rush, Malkan \& Spinoglio (1993), converted to 15 $\mu$m using appropriate SEDs
(41 type 1 AGNs and 50 type 2 AGNs).   
The typical SED assumed for type 1 AGNs is that compiled by Elvis et al. (1994), 
while for type 2 AGN two extreme cases of the obscured AGN SED in the MIR
were assumed: a starburst-like SED (Circinus; Sturm et al. 2000) and an AGN-like
SED (NGC 1068; Sturm et al. 2000). 
Type 1 AGNs are found to evolve 
with a luminosity evolution [$L(z) \propto (1+z)^{k_L}$)], with an evolution 
rate $k_L$ equal to 2.6 up to $z \sim 2$ and constant thereon.
A similar evolutionary scenario is found for type 2 AGNs, with $k_L$ ranging from 2.0 to 2.6
depending on the adopted SED (NGC 1068 or Circinus, respectively).
The best-fitting model is found to reproduce well both observed source counts and
redshift distributions, as shown by I. Matute et al. (2005, in preparation).

\section{Counts at 24 micron}

\subsection{Model Predictions at 24 $\mu$m}
\label{exp24mic}
\begin{figure}
\plotone{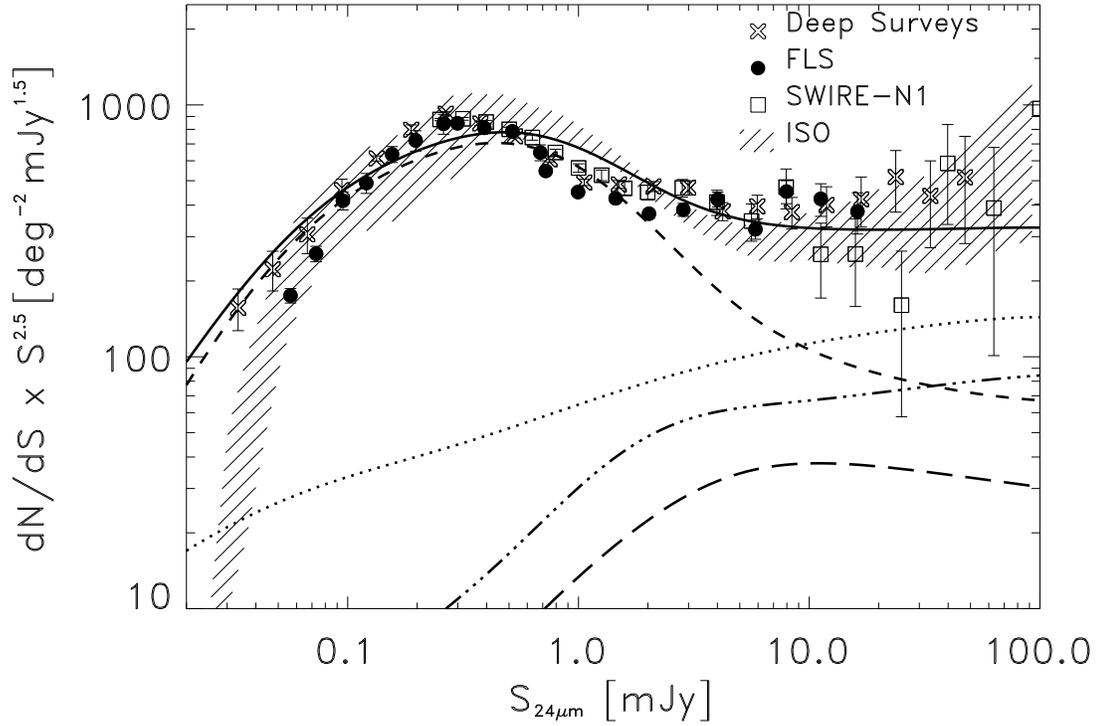}
\caption{Normalised differential source counts at 24 $\mu$m (crosses: Deep
{\em Spitzer} Surveys, Papovich et al. 2004; filled circles: FLS, Marleau et al.
2004; open squares: SWIRE-N1, Shupe et al. in prep.) with the model predictions 
(as in figure \ref{fig2}) and the 15 \mic data (shaded region)
transformed to 24 \mic as described in the text.}
\label{fig3}
\end{figure}

By using the appropriate SED for each population and convolving the
  SED with the appropriate filter transmission, the galaxy and AGN local LFs have been transformed 
from 15 to 24 $\mu$m. Then the predicted 24 $\mu$m source counts have 
been computed for all the contributing populations. In figure \ref{fig3}, the source counts
predicted by our model are compared to the recently published {\em Spitzer} 
24 $\mu$m data from the FLS (Marleau et al. 2004),
  from the deep surveys (Papovich et al. 2004) and from the {\em Spitzer} Wide-Area Infra-red
Extragalactic Survey (SWIRE; D.L. Shupe et al. 2005, in preparation). We can notice a consistency
between data and model, with no need, at least in first approximation, for the use
of more extreme SEDs for starburst galaxies (i.e., Arp 220). 
AGNs (either type 1 or 2) do not dominate
the observed source counts at any flux level, although type 2 make about 25\% of the counts 
at $S_{24 \mu m} > 10$ mJy. The counts are dominated by non evolving normal galaxies at 
$\gsimeq 8$ mJy and by evolving starburst galaxies at lower flux densities.
The assumption made in our model of no evolution for galaxies at $z > 1$, not very well
constrained by ISOCAM data, is ``a posteriori'' consistent with {\em Spitzer}
data.

\begin{figure}
\plotone{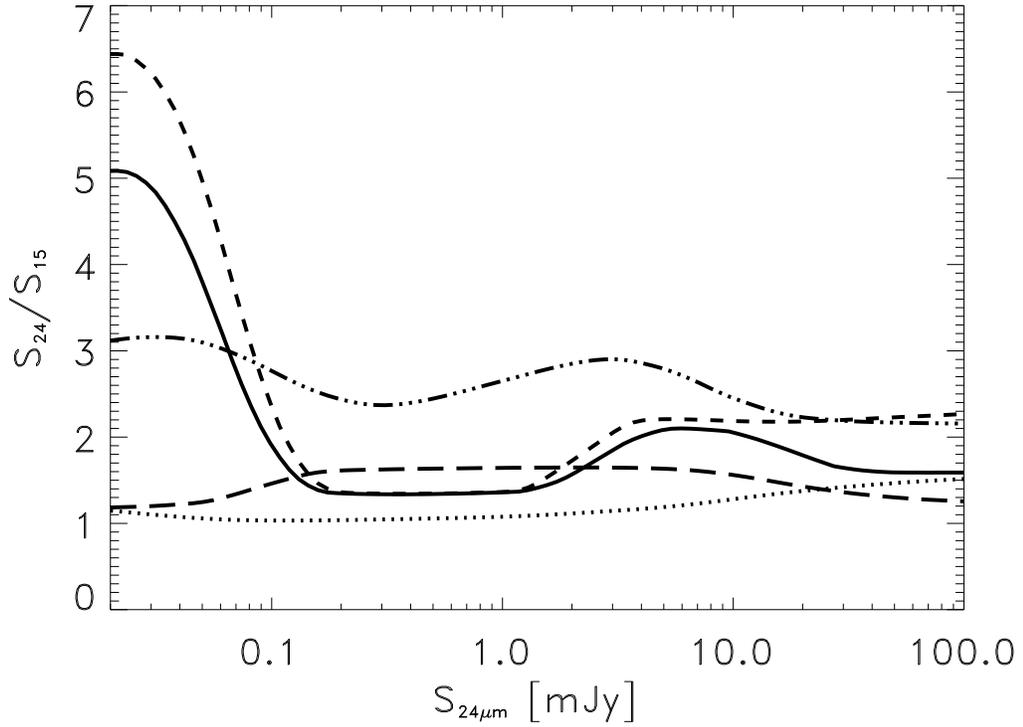}
\caption{Median values of the $S_{24 \mu m} / S_{15 \mu m}$ flux density ratio 
as function of $S_{24 \mu m}$, as derived by our model for all the populations 
(weighted mean: solid line; different populations: lines as in previous figures).
}
\label{fig4}
\end{figure}

\begin{figure}
\plotone{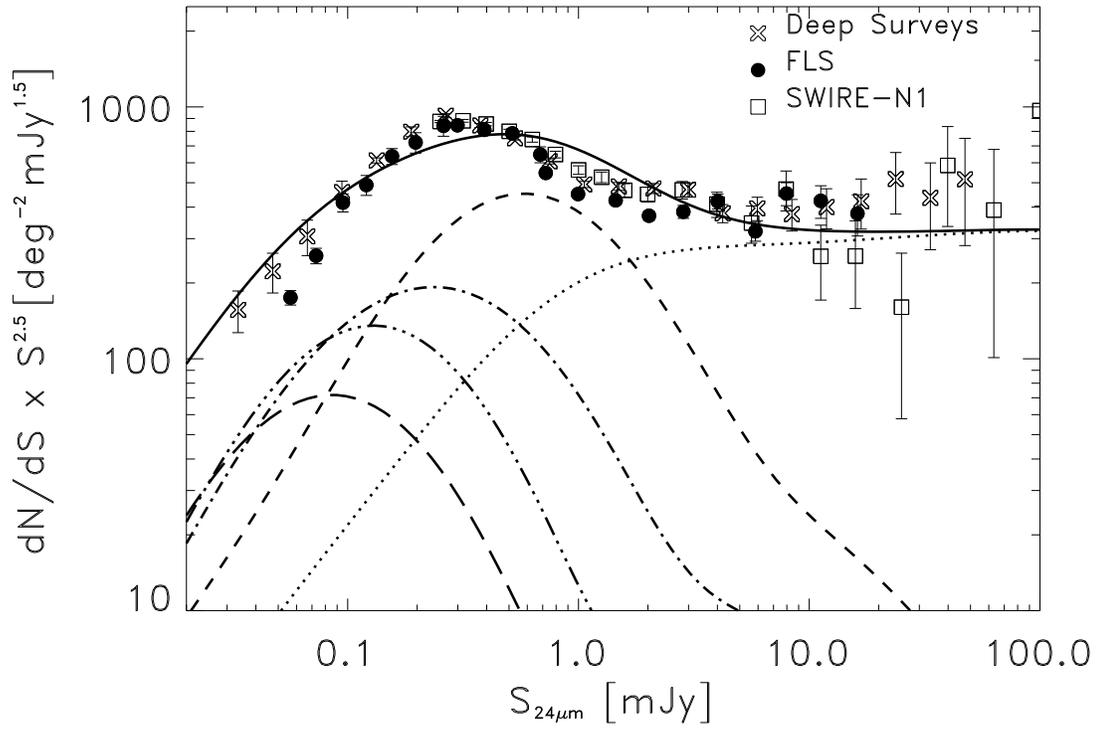}
\caption{Differential redshift contribution to the normalized differential source
  counts at 24 $\mu$m. The dotted, dashed, dot-dashed, dot-dot-dot-dashed
and long-dashed lines correspond to the 0.0--0.5, 0.5--1.0, 1.0--1.5, 1.5--2.0
and 2.0--2.5 redshift intervals respectively.}
\label{fig5}
\end{figure}

It is interesting to see how
the ratio between the {\em Spitzer} 24  and the ISOCAM 15 $\mu$m flux for all
the populations changes as a function of $z$ (figure \ref{fig1}) and 
of the 24 $\mu$m flux (as derived by our model; figure \ref{fig4}).
The comparison between figs \ref{fig1} and \ref{fig4} clearly shows that
the higher flux densities ($S_{24 \mu m} > 2-3$ mJy) are dominated by nearby objects with  
moderately high values of the $S_{24 \mu m} / S_{15 \mu m}$ ratio ($\sim 2.2$: starburst 
and type 2 AGNs; $\sim 1.7$: normal galaxies), while the bump of the 24 \mic  
counts (at fluxes 0.1 -- 1 mJy) is dominated
by objects with $S_{24 \mu m} / S_{15 \mu m} \approx 1.4$ (mainly starburst galaxies
at $0.7 \lsimeq z \lsimeq 1.5$). These are
the same populations found to contribute to the ISOCAM 15 $\mu$m source counts. However, since
we have approximately
\begin{eqnarray}
\left( \frac{dN}{dS} S^{2.5} \right)_{24 \mu m} = \left( \frac{dN}{dS} S^{2.5} \right)_{15 \mu m} \left(
\frac{S_{24 \mu m}}{S_{15 \mu m}} \right)^{1.5}.
\end{eqnarray}
and the ratio value at flux densities around the peak of the differential source counts
is $\simeq 1.4$, the evolutionary excess is more pronounced at 15 $\mu$m than at 24 $\mu$m.
At the lower flux densities ($S_{24 \mu m} \lsimeq 0.1$ mJy), sources with high flux ratio values
($S_{24 \mu m} / S_{15 \mu m} > 2.0 - 2.5$, corresponding to $z > 1.5$) start to 
dominate the counts. 
These high-$z$ starburst galaxies, mainly at $1.5 \leq z \leq 2.5$, are not visible in the 
ISOCAM surveys, because of the 15 $\mu$m $k$-correction,
but contribute to the fainter end of the 24 $\mu$m source counts and
to the 24 $\mu$m cosmic background. Since the existence of this starburst population at high 
redshift is predicted just by extrapolating our 15 \mic model to higher $z$, it is likely that 
faint 24 \mic sources are the high-redshift likes of the fainter 15 \mic sources detected by ISOCAM  
(see also Chary et al. 2004).\\
The same kind of considerations are evident from figure \ref{fig5}, where the 
contributions to the 24 \mic number counts from different redshift intervals are shown. 
While galaxies with $0.0 \lsimeq z \lsimeq 0.5$ dominate the high fluxes ($S_{24 \mu m}\gsimeq{2}$ mJy), 
the peak of the differential counts is made manly by $0.5 \lsimeq z \lsimeq 1.5$ sources, although
at $S_{24 \mu m}\lsimeq{0.16}$ mJy (corresponding roughly to the limit of the deepest, non-lensed,
ISOCAM survey: $S_{15 \mu m} = 0.12$ mJy; see figure \ref{fig5}) the higher-$z$ population 
contributes for a conspicuous fraction ($\sim$46\%).
By extrapolating the source counts model down to the faintest 
fluxes, we obtain an estimate of the total 24 \mic background:
${\nu}I_{\nu}(24{\mu}m){\sim}2.4$ nWm$^{-2}$sr$^{-1}$. According to our model, the
amount produced by sources at $z\lsimeq$1.5 is ${\sim}1.6$ nWm$^{-2}$sr$^{-1}$,
therefore $\sim$66\% of the 24 $\mu$m background originates at relatively low redshift.
From the comparison between the total background predicted by our model and the value
derived from the observed 24 \mic
source counts ($1.9\pm{0.6}$ nW m$^{-2}$ sr$^{-1}$; Papovich et
al. 2004), we find that the {\em Spitzer} deep surveys already resolve
$\sim$78\% of the total 24 $\mu$m background.

\subsection{Observed ISOCAM Counts Transformed to 24 $\mu$m}
\label{15mic_24mic}

To compare observed data counts at 15 \mic with those at 24 $\mu$m, we have converted 
the 15 \mic source counts obtained from different ISOCAM surveys (from the
ultra-deep lensed of Metcalfe et al. 2003, to
the shallower ELAIS Survey of Gruppioni et al. 2002) to 24 $\mu$m, as described
in the following text.
We have convolved the observed 15 \mic differential counts ($\frac{dN}{dS_{15 \mu m}}$)
with the distribution of
the ratios $S_{24 \mu m}/S_{15 \mu m}$ obtained from our model, given a 15 \mic
source selection ($f[(S_{24 \mu m}/S_{15 \mu m}),S_{15 \mu m}]$):
\begin{equation}
dN(S_{24 \mu m})=\int{f\left(\frac{S_{24 \mu m}}{S_{15 \mu m}},S_{15 \mu m}\right)\frac{dN}{dS_{15 \mu m}}(S_{15 \mu m})dS_{15 \mu m}}
\end{equation}
\noindent 
Data at 15 \mic from different samples have been combined by weighting
each point by its formal error (inverse of the squared error).
In figure \ref{fig3} the counts derived from
the 15 \mic observed data (shaded area) are 
overplotted to the 24 \mic data and model.
The two different source counts are consistent and, in first approximation, seem to
agree well. In particular, at
high flux densities we do not observe the discrepancy as
reported by Marleau et al. (2004), thanks to a
more accurate flux density ratio applied. 
On the other hand, the discrepancy observed at 24 \mic fluxes lower than $\sim$0.05 mJy
is only apparent, since fluxes fainter than this are not
sampled by the 15 \mic data. Some level of
inconsistency between the two source counts are visible around $\sim$1 mJy, where the 15
$\mu$m counts (and the model) are higher than the
observed 24 \mic source counts. A decrease of the $S_{24 \mu m}/S_{15 \mu m}$ ratio 
around 1 mJy could be obtained by slightly modifying the
starburst template in the MIR range (i.e. increasing the PAH features 
with respect to the continuum, since at the typical redshift of sources with
these flux densities, $z \approx 1$, the PAH features enter the 15 \mic band).
A similar change was made by Lagache et al. (2004), who show how
a minor change in the starburst template spectra between 12 and 30 \mic (together
with a slight modification of the luminosity density) was a sufficient a posteriori
modification to enable a model not fitting the observed 24 \mic source counts (Lagache et al. 2003) 
to reproduce the observations.\\ 
We are actually working at improving the model--data agreement by including also
the recently published 24 $\mu$m source counts (and all the MIR-FIR counts 
and redshift distributions available in literature) as an observational constraint
to the maximum likelihood analysis of our 15 $\mu$m-based model. Moreover, we
are considering the use of an SED library, with different SEDs associated not only
to different populations, but also to different luminosity classes (i.e. Chary \&
Elbaz 2001). The model improvements will be described in a future
paper (F. Pozzi et al. 2005, in preparation), since they are beyond the scope of the
present Letter, whose intent is just to show, through a simple 15 $\mu$m-based
model, what the 24 $\mu$m/15 $\mu$m comparison can tell us in terms of galaxy
evolutionary properties.

\section{Conclusions}
\label{concl}
We have discussed the importance of a comparison between the extragalactic
source counts in two different MIR bands for the study of the evolutionary
properties of galaxies. We have shown what we expect in terms of
variations of the $S_{24 \mu m} / S_{15 \mu m}$ flux density ratio
with redshift and 24 \mic flux, considering typical SEDs for the
different populations contributing to the source counts.
Then, to compare the observed {\em Spitzer} 24 \mic source counts with the 
ISOCAM 15 \mic ones, we have presented the prediction in the 
24 \mic {\em Spitzer} band from a phenomenological 
evolution model based on the ISOCAM 15 \mic LF of 
galaxies and AGNs. Actually, this model is 
the only one available in literature
that is able to reproduce the observed 24 \mic source counts without
the need, at least in first approximation, of any a posteriori updates. 
We have also shown that the
observed ISOCAM data points transformed from 15 \mic to 24 \mic seem
to agree well with the recently published {\em Spitzer} source counts.
Our model suggests the appearance of a new population of high redshift ($z > 1.5$) galaxies
at 24 $\mu$m, not detected in the previous ISOCAM surveys, but probably
the high-$z$ likes of the 15 \mic sources.

\acknowledgments
We thank the anonymous referee for valuable comments that improved
the quality of this Letter and D.L. Shupe and I. Matute for kindly providing
SWIRE data counts and AGN models before publication.


\clearpage


\end{document}